\documentclass[11pt,a4paper]{article}
\usepackage{amssymb}

\makeatletter

\def\Reals{\mathbb{R}}
\def\Integers{\mathbb{Z}}
\def\suchthat{\,:\,}
\def\of{{{}_\circ}}
\def\ox{\otimes}

\def\ket#1{{\left|#1\right\rangle}}

\def\proj#1{{\left|#1\right\rangle\!\left\langle#1\right|}}
\def\braket#1#2{{\left\langle#1\vphantom{#2}\right|\left.\vphantom{#1}#2\right\rangle}}

\newtheorem{theorem}{Theorem}
\def\quoderat{\hbox{\vbox{\hrule height\fboxrule\hbox{\vrule width\fboxrule\vbox{\vskip 2.6mm}\hskip2.6mm\vrule width\fboxrule}\hrule height\fboxrule}}}
\newenvironment{proof}
 {\vspace{13pt}\noindent{\bf Proof}\hspace{0.7\parindent}}
 {\hfill\quoderat\vspace{13pt}}

\makeatother

\begin{document}

\title{Quantum computational gradient estimation}
\author{David W. Bulger\thanks{Lecturer, Department of Statistics, Macquarie University, NSW 2109, Australia.
                 {\tt dbulger@efs.mq.edu.au}.}}

\maketitle

\thispagestyle{empty}

\begin{abstract}
Classically, determining the gradient of a black-box function $f:\Reals^p\to\Reals$ requires $p+1$ evaluations.
Using the quantum Fourier transform, two evaluations suffice.
This is based on the approximate local periodicity of $e^{2\pi i\lambda f(x)}$.
It is shown that sufficiently precise machine arithmetic results in gradient estimates of any required accuracy.
\end{abstract}

\noindent{\bf MSC2000 Subject Classification:}\ 90C30, 68Q99, 68Q25\\[0mm]

\section{Introduction}

The vector gradient of a real-valued function $f$ of a vector argument can be calculated
using just two calls to a black-box quantum oracle for $f$.
The mechanism is simple, and capitalises on the fact that, in the vicinity of a point $x$,
$e^{2\pi i\lambda f(x)}$ is periodic, with period parallel and inversely proportional to $\nabla f(x)$.
A superposed state is created discretising a small hyperrectangle around the domain point, the function is evaluated,
the phase is rotated in proportion to the function value, the oracle call is reversed,
and a multidimensional quantum Fourier transform is applied to the bits encoding the discretised hyperrectangle.

This paper establishes, under mild conditions on $f$,
that the gradient estimation can be performed to any required level of accuracy,
in the sense that, given any $\delta>0$ and $\epsilon<1$, we can produce a superposition of gradient estimates which,
if observed, will collapse to an estimate within $\delta$ of the true gradient with probability at least $\epsilon$.
Greater accuracy is achieved by increasing arithmetic precision and by increasing the number of points in the sampling grid.
 
The paper's structure is as follows.
Section~\ref{secProblem} presents some assumptions on the function $f$ whose gradient is sought,
and Section~\ref{secOracle} formalises the evaluation and manipulation of values of $f$ within the quantum computer.
Section~\ref{secAlg} presents the gradient estimation algorithm.
Section~\ref{secBehaviour} analyses the effect of the algorithm,
consisting mostly of the statement and proof of the main result,
that any required accuracy is attainable by using sufficiently precise arithmetic and a large enough sampling grid.
Because the rest of paper discusses the computations rather abstractly,
Section~\ref{secImplementation} briefly comments on how the algorithm would be performed in practice.

\section{Problem formulation}  \label{secProblem}

Let $D\subset\Reals^p$ have non-empty interior.
Let $f$ be a twice-differentiable function from $D$ to $\Reals$.
At each point $x\in D$, let $\nabla f(x)$ denote the gradient and $Hf(x)$ the Hessian matrix of $f$.
Assume that $\|\nabla f(x)\|_\infty\leq L$ and $\|Hf(x)\|_2\leq M$ for all $x\in D$.
It is desired to determine $\nabla f(x)$ for a point $x$ in the interior of $D$.

\section{Oracle Formalism}  \label{secOracle}

This paper's main result is that an objective function's gradient can be calculated to any desired precision using the quantum algorithm described.
Clearly, any particular data encoding method will only support a certain maximum precision;
we therefore require a formalism in which points in the domain and range of $f$ can be represented in a variety of ways.
This section introduces the Hilbert spaces and operators involved.

It will be helpful firstly to catalog the operators to be used as they would look
if precision and rounding error were {\em not\/} relevant.
The computational system is a tripartite system; the three parts have state spaces
$\cal D$, $\cal R$ and $\cal G$ (standing for `domain', `range' and `grid'),
so the combined system has state space $\cal D\ox R\ox G$.
Each computational basis state is a tensor product $\ket d\ox\ket r\ox\ket g$
of one computational basis state from each of the three factor spaces.
For now, suppose that the basis indices $d$, $r$ and $g$ belong respectively to $D$, $\Reals$ and $\Reals^p$.
Let $g_0$ be the constant length-$p$ vector $(2^{n-1}-1/2,\ldots,2^{n-1}-1/2)$.
The operators are
\begin{eqnarray*}
U_f(\ket d\ox\ket r\ox\ket g) & = & \ket d\ox\ket{r+f(d)}\ox\ket g, \\
U_+(\ket d\ox\ket r\ox\ket g) & = & \ket{d+\mu(g-g_0)}\ox\ket r\ox\ket g, \\
U_R(\ket d\ox\ket r\ox\ket g) & = & e^{2\pi i\lambda f(d)}\ket d\ox\ket r\ox\ket g,
\end{eqnarray*}
where $\mu$ and $\lambda$ are algorithm parameters,
as well as the quantum Fourier transform and inverses of $U_f$ and $U_+$.

Return now to precision considerations.
Suppose that, for any positive $\nu$ and $\mu$ and natural $n$,
we can construct a quantum oracle evaluating $f(x+\mu(g-g_0))$ for $g\in\{0,\ldots,2^n-1\}^p$ and $x,x+\mu(g-g_0)\in D$,
with an error uniformly bounded by $\nu$.
In particular, suppose that we can construct a system
$(U_f, U_+, {\cal D}, B_{\cal D}, {\cal R}, B_{\cal R}, {\cal G}, c_d, c_r, c_f, c_p)$, where
\begin{itemize}
\item ${\cal D}$, $\cal R$ and ${\cal G}$ are finite-dimensional Hilbert spaces,
\item $B_{\cal D}$ and $B_{\cal R}$ are orthonormal bases for ${\cal D}$ and ${\cal R}$,
\item $B_{\cal G}=\{0,\ldots,2^n-1\}^p$ is an orthonormal basis for $\cal G$,
\item $B_{\cal R}$ is a group, under an operation we denote as `$+$',
 with an identity element we denote as $\ket0$,
\item $c_d:D\to B_{\cal D}$ (the `domain encoding function'), $c_f:B_{\cal D}\to B_{\cal R}$,
 $c_r:B_{\cal R}\to\Reals$ (the `range decoding function'), and $c_p:B_{\cal D}\times B_{\cal G}\to B_{\cal D}$,
\item $U_f$ and $U_+$ are unitary operators on $\cal D\ox R\ox G$, given by
 \begin{eqnarray*}
 U_f\ket d\ox\ket r\ox\ket g & = & \ket d\ox\ket{r+c_f(d)}\ox\ket g, \\
 U_+\ket d\ox\ket r\ox\ket g & = & \ket{c_p(d,g)}\ox\ket r\ox\ket g,
 \end{eqnarray*}
\item $c_p$ acts invertibly on $B_{\cal D}$, so that for each $d\in B_{\cal D}$ and $g\in B_{\cal G}$,
$c_p^{-1}(c_p(d, g), g)=d$,
\item $|f(x)-c_r\of c_f\of c_d(x)|\leq\nu/2$ for all $x\in D$,
\item $|c_r\of c_f\of c_d(x+\mu(g-g_0))-c_r\of c_f\of c_p(c_d(x),g)|\leq\nu/2$ for all $x\in D$ and $g\in B_{\cal G}$,
 provided $x+\mu(g-g_0)$ is also in $D$,
\end{itemize}
and, further, that we can implement $U_f$ and $U_+$ on a quantum computer.
(Note that this formalism does not necessarily require the domain points represented by $B_{\cal D}$ to form a grid;
this may be of interest in optimising functions on manifolds.)

\section{Algorithm}  \label{secAlg}

The algorithm dealt with in this paper estimates the gradient of $f$ at a point $x$ in the interior of its domain.
Firstly, using quantum superposition,
$f$ is evaluated at every point of a hyperrectangular grid centred around $x$.
The grid is small enough that $f$ is approximately linear across it.
Next, the phase of the quantum computational system is rotated in proportion to the value of $f$ at each grid point.
Now the phase varies approximately periodically over the grid, and the period determines $\nabla f(x)$.
The period is easily determined by the quantum Fourier transform.

Two of the operators involved in the gradient estimation algorithm, $U_f$ and $U_+$,
were hypothesised in Section~\ref{secOracle}.
Additionally, we will require their inverses $U_f^{-1}$ and $U_+^{-1}$, a phase rotation operator $U_R$,
and a $p$-dimensional quantum Fourier transform $U_{QFT}$.

The operators $U_f^{-1}$ and $U_+^{-1}$ invert the actions of $U_f$ and $U_+$, mapping $\ket d\ox\ket r\ox\ket g$ to
$\ket d\ox\ket{r-c_f(d)}\ox\ket g$ (the subtraction $r-c_f(d)$ is according to the group structure assumed on $B_{\cal R}$)
and $\ket{c_p^{-1}(d,g)}\ox\ket r\ox\ket g$.
Note that the function $f$ is {\em not\/} being inverted.

The phase rotation operator $U_R$ involves a parameter $\lambda\in\Reals$, mapping $\ket d\ox\ket r\ox\ket g$ to
$e^{2\pi i\lambda c_r(r)}\ket d\ox\ket r\ox\ket g$.
The multidimensional quantum Fourier transform $U_{QFT}$ acts on $\cal G$, mapping $\ket d\ox\ket r\ox\ket g$ to
$2^{-pn/2}\sum_{h\in B_{\cal G}}e^{2\pi ih\cdot g/2^n}\ket d\ox\ket r\ox\ket h$.

With these operators defined, the gradient estimation algorithm is easily stated.
Firstly, the state $\ket{c_d(x)}\ox\ket0\ox\ket0$ is prepared in $\cal D\ox R\ox G$,
where $x\in D$ is the point at which $\nabla f$ is sought.
Then, the system $\cal D\ox R\ox G$ is subjected to $U_{QFT}\of U_+^{-1}\of U_f^{-1}\of U_R\of U_f\of U_+\of U_{QFT}$.
This results, as we shall see in Section~\ref{secBehaviour}, in a state $\ket{c_d(x)}\ox\ket0\ox\ket\chi$,
where in general $\ket\chi$ is a superposition of computational basis states from $B_{\cal G}$.

Interpretation of the resulting state $\ket\chi$ involves the ``gradient decoding function'' $c_g:B_{\cal G}\to\Reals^p$,
defined by
\begin{eqnarray*}
c_g:(g_1, \ldots, g_p) & \to & (c_{g,1}(g_1), \ldots, c_{g,p}(g_p)), \mbox{ where} \\
c_{g,m}(g_m) & = & \left\{\begin{array}{ll} \frac{-g_m}{2^n\lambda\mu}, & g_m\in\{0, \ldots, 2^{n-1}-1\}, \\
\frac{2^n-g_m}{2^n\lambda\mu}, & g_m\in\{2^{n-1}, \ldots, 2^n-1\}.\end{array}\right.
\end{eqnarray*}
If $\ket\chi$ is a basis state $\ket g$, it indicates that $\nabla f(x)=c_g(g)$.
If, on the other hand, $\ket\chi$ is a superposition $\sum_{g\in B_{\cal G}}\chi_g\ket g$,
then the gradient estimate is indeterminate, comprising the various discretised values $g$ with the weights $|\chi_g|^2$.

Altogether, in addition to the argument $x$, the gradient estimation algorithm depends on the four parameters $n$, $\nu$, $\lambda$ and $\mu$.
Accordingly, the algorithm will be denoted $A(n, \nu, \lambda, \mu; x)$.

\section{Behaviour}  \label{secBehaviour}

The state resulting from the algorithm $A(n, \nu, \lambda, \mu; x)$ is
\begin{eqnarray*}
\lefteqn{U_{QFT}\of U_+^{-1}\of U_f^{-1}\of U_R\of U_f\of U_+\of U_{QFT}(\ket{c_d(x)}\ox\ket0\ox\ket0)} \\
& = & 2^{-pn/2}\sum_{h\in B_{\cal G}}U_{QFT}\of U_+^{-1}\of U_f^{-1}\of U_R\of U_f\of U_+(\ket{c_d(x)}\ox\ket0\ox\ket h) \\
& = & 2^{-pn/2}\sum_{h\in B_{\cal G}}U_{QFT}\of U_+^{-1}\of U_f^{-1}\of U_R\of U_f(\ket{c_p(c_d(x),h)}\ox\ket0\ox\ket h) \\
& = & 2^{-pn/2}\sum_{h\in B_{\cal G}}U_{QFT}\of U_+^{-1}\of U_f^{-1}\of U_R(\ket{c_p(c_d(x),h)}\ox\ket{c_f\of c_p(c_d(x),h)}\ox\ket h) \\
& = & 2^{-pn/2}\sum_{h\in B_{\cal G}}e^{2\pi i\lambda c_r\of c_f\of c_p(c_d(x),h)}U_{QFT}\of U_+^{-1}\of U_f^{-1}(\ket{c_p(c_d(x),h)}\ox\ket{c_f\of c_p(c_d(x),h)}\ox\ket h) \\
& = & 2^{-pn/2}\sum_{h\in B_{\cal G}}e^{2\pi i\lambda c_r\of c_f\of c_p(c_d(x),h)}U_{QFT}\of U_+^{-1}(\ket{c_p(c_d(x),h)}\ox\ket0\ox\ket h) \\
& = & U_{QFT}\ket{c_d(x)}\ox\ket0\ket\psi \\
& = & \ket{c_d(x)}\ox\ket0\ox\ket\chi,
\end{eqnarray*}
where
\begin{eqnarray}
\ket\psi & = & 2^{-pn/2}\sum_{h\in B_{\cal G}}e^{2\pi i\lambda c_r\of c_f\of c_p(c_d(x),h)}\ket h \mbox{ and} \\
\ket\chi & = & U_{QFT}\ket\psi,  
\end{eqnarray}
and of course $U_{QFT}$, acting on $\cal G$ alone, is defined by
$$
U_{QFT}(\ket g) = 2^{-pn/2}\sum_{h\in B_{\cal G}}e^{2\pi ih\cdot g/2^n}\ket h.
$$

\begin{theorem}  \label{thmCgnce}
For any $\gamma,\delta>0$ and $\epsilon<1$,
there exist parameters $n$, $\nu$, $\lambda$ and $\mu$ such that,
at every $x$ with $x+[-\gamma,\gamma]^d\subseteq D$,
when $\ket\chi$ is produced according to $A(n, \nu, \lambda, \mu; x)$,
\begin{equation}  \label{ineqTarget}
\left\|P\ket\chi\right\|_2\geq\epsilon,
\end{equation}
where $P$ is the projection
$$
\sum\left\{\proj h\suchthat\|c_g(h)-\nabla f(x)\|_\infty<\delta\right\}.
$$
\end{theorem}

\begin{proof}
It will be demonstrated that, if $n$, $\nu$, $\lambda$ and $\mu$ are chosen satisfying
\begin{eqnarray}
4^{n-1}\pi\lambda M\mu^2/\sqrt5 & \leq & (1-\epsilon)/3, \label{ineqNonlin} \\
2\pi\lambda\nu & \leq & (1-\epsilon)/3, \label{ineqPrecision} \\
2^{n-1}\mu & \leq & \gamma, \label{ineqMargin} \\
1/2\lambda\mu & \geq & L+\delta \mbox{ and} \label{ineqBandwidth} \\
\csc(\pi\lambda\mu\delta) & \leq & \sqrt{2^n(1-((2+\epsilon)/3)^{2/p})}, \label{ineqLeakage}
\end{eqnarray}
then~(\ref{ineqTarget}) holds.
The reader can verify that one choice satisfying~(\ref{ineqNonlin}) to~(\ref{ineqLeakage}) is
\begin{eqnarray*}
n & = & \left\lceil-\log_2\left(\sin^2(\pi\delta/2(L+\delta))\left(1-((2+\epsilon)/3)^{2/p}\right)\right)\right\rceil, \\
\lambda & = & \max\left\{\frac{2^{n-2}}{\gamma(L+\delta)}, \frac{3\times4^{n-2}\pi M}{\sqrt5(L+\delta)^2(1-\epsilon)}\right\}, \\
\mu & = & 1/2\lambda(L+\delta), \\
\nu & = & (1-\epsilon)/6\pi\lambda.
\end{eqnarray*}

The algorithm contains three sources of error:
\begin{itemize}
\item $\nabla f(x)$ will not, in general, be exactly equal to $c_g(g)$ for some $g\in B_{\cal G}$;
\item $\nabla f$ will not, in general, be exactly constant throughout the sampling grid;
\item calculations are performed to a finite precision, so that $c_r\of c_f\of c_p(c_d(x),h)$ will not,
in general, exactly equal $f(x+\mu(h-g_0))$.
\end{itemize}

In the oracle's calculation of $f(x+\mu(h-g_0))$, let $\epsilon_D(x,h)$ represent the error due to computational precision
and let $\epsilon_N(x, h)$ represent the departure from linearity, so that
$$
c_r\of c_f\of c_p(c_d(x),h) = f(x+\mu(h-g_0)) + \epsilon_D(x, h) = f(x) + \nabla f(x)\cdot\mu(h-g_0) + \epsilon_N(x, h) + \epsilon_D(x, h).
$$
(`D' stands for `discretisation' and `N' for `nonlinear'.)
By~(\ref{ineqMargin}), $x+\mu(h-g_0)\in D$.
By assumption, $|\epsilon_D(x, h)|\leq\nu$, and by Lagrange's remainder for Taylor's series,
we have $|\epsilon_N(x, h)|\leq M\mu^2(h-g_0)\cdot(h-g_0)/2$.

We wish to bound the effects of the three sources of error separately;
therefore it will be convenient to write $\ket\psi$ as the sum $\ket{\psi_L}+\ket{\psi_N}+\ket{\psi_D}$,
where
\begin{eqnarray*}
\ket{\psi_L} & = & 2^{-pn/2}\sum_{h\in B_{\cal G}}e^{2\pi i\lambda(f(x) + \mu\nabla f(x)\cdot(h-g_0))}\ket h, \\
\ket{\psi_N} & = & 2^{-pn/2}\sum_{h\in B_{\cal G}}\left[
 e^{2\pi i\lambda f(x+\mu(h-g_0))}-e^{2\pi i\lambda(f(x)+\mu\nabla f(x)\cdot(h-g_0))}\right]\ket h \\
& = & 2^{-pn/2}\sum_{h\in B_{\cal G}}\left[
 e^{2\pi i\lambda f(x+\mu(h-g_0))}-e^{2\pi i\lambda(f(x+\mu(h-g_0))-\epsilon_N(x,h))}\right]\ket h, \\
\ket{\psi_D} & = & 2^{-pn/2}\sum_{h\in B_{\cal G}}\left[
 e^{2\pi i\lambda c_r\of c_f\of c_p(c_d(x),h)} - e^{2\pi i\lambda f(x+\mu(h-g_0))}\right]\ket h \\
& = & 2^{-pn/2}\sum_{h\in B_{\cal G}}\left[
 e^{2\pi i\lambda(f(x+\mu(h-g_0))+\epsilon_D(x,h))}-e^{2\pi i\lambda f(x+\mu(h-g_0))}\right]\ket h. \\
\end{eqnarray*}
Noting that, for any real $\alpha$ and $\beta$,
$$
|e^{2\pi i\lambda(\alpha+\beta)} - e^{2\pi i\lambda\alpha}| = 2|\sin\pi\lambda\beta| \leq 2\pi\lambda|\beta|,
$$
we have
\begin{eqnarray*}
\|\ket{\psi_N}\|_2
& \leq & 2^{-pn/2}\sqrt{\sum_{h\in B_{\cal G}}(\pi\lambda M\mu^2(h-g_0)\cdot(h-g_0))^2} \\
& = & 2^{-pn/2}\pi\lambda M\mu^2\sqrt{2^{dn}\left(
 \frac{16^n}{80}-\frac{4^n}{24}+\frac{2^n}{12}+\frac7{240}-\frac1{12\times2^n}\right)} \\
& \leq & 4^{n-1}\pi\lambda M\mu^2/\sqrt{5} \\
& \leq & (1-\epsilon)/3
\end{eqnarray*}
by~(\ref{ineqNonlin}), and
$$
\|\ket{\psi_D}\|_2
\leq 2^{-pn/2}\sqrt{\sum_{h\in B_{\cal G}}(2\pi\lambda\nu)^2}
= 2^{-pn/2}\times2\pi\lambda\nu\sqrt{|B_{\cal G}|}
= 2\pi\lambda\nu \\
\leq (1-\epsilon)/3
$$
by~(\ref{ineqPrecision}).

Next we consider the error introduced by `frequency leakage'.
If the components of $\nabla f(x)$ are integer multiples of $1/2^n\lambda\mu$,
then $U_{QFT}\ket{\psi_L}$ is equal to a computational basis state,
identifying $\nabla f(x)$ exactly.
In the general case, we obtain instead a superposition,
which strongly weights computational basis states representing gradients close to $\nabla f(x)$.
We have
\begin{eqnarray*}
U_{QFT}\ket{\psi_L} & = & 2^{-pn}\sum_{g\in B_{\cal G}}\sum_{h\in B_{\cal G}}
 e^{2\pi i\left(g\cdot h/2^n+\lambda(f(x)+\mu\nabla f(x)\cdot(h-g_0))\right)}\ket j \\
& = & e^{2\pi i\lambda(f(x)-\mu\nabla f(x)\cdot g_0)}\bigotimes_{m=1}^p\ket{\phi_m}, \mbox{ where} \\
\ket{\phi_m} & = & 2^{-n}\sum_{g_m=0}^{2^n-1}\sum_{h_m=0}^{2^n-1}
 e^{2\pi ih_m\left(\frac{g_m}{2^n}+\lambda\mu\frac{\partial f(x)}{\partial x_m}\right)}\ket{g_m} \\
& = & 2^{-n}\sum_{g_m=0}^{2^n-1}\frac
 {1-e^{2\pi i\left(g_m+2^n\lambda\mu\frac{\partial f(x)}{\partial x_m}\right)}}
 {1-e^{2\pi i\left(\frac{g_m}{2^n}+\lambda\mu\frac{\partial f(x)}{\partial x_m}\right)}}\ket{g_m}.
\end{eqnarray*}
The factors $\ket{\phi_m}$ are state vectors of unit magnitude, and note that
\begin{eqnarray}
|\braket{g_m}{\phi_m}| & \leq & \frac{2^{1-n}}{\left|e^{\pi i\left(\frac{g_m}{2^n}+\lambda\mu\frac{\partial f(x)}{\partial x_m}\right)}
- e^{-\pi i\left(\frac{g_m}{2^n}+\lambda\mu\frac{\partial f(x)}{\partial x_m}\right)}\right|} \nonumber \\
& = & 2^{-n}\left|\csc\left(\pi\left(\frac{g_m}{2^n}+\lambda\mu\frac{\partial f(x)}{\partial x_m}\right)\right)\right| \nonumber \\
& = & 2^{-n}\left|\csc\left(\pi\lambda\mu\left(c_{g,m}(g_m)-\frac{\partial f(x)}{\partial x_m}\right)\right)\right| \label{eqCosecant} \\
& \leq & 2^{-n}|\csc(\pi\lambda\mu\delta)| \label{ineqCosecant}
\end{eqnarray}
whenever $|c_{g,m}(h_m)-\partial f(x)/\partial x_m|\geq\delta$;
we obtain~(\ref{eqCosecant}) because $\lambda\mu c_{g,m}(g_m)+g_m/2^n$ is always an integer,
and $|\csc|$ is even and has period $\pi$;
we obtain~(\ref{ineqCosecant}) due to the shape of the cosecant function and because, by~(\ref{ineqMargin}),
$$
\pi\lambda\mu(c_{g,m}(g_m)-\partial f(x)/x_m) \in (-\pi+\pi\lambda\mu\delta,\pi-\pi\lambda\mu\delta).
$$

Note that the projection $P$ can be written as $P_1\ox\cdots\ox P_d$, where
$$
P_m = \sum_{h_m=0}^{2^n-1}\left\{\proj{h_m}\suchthat
 \left|c_{g,m}(h_m)-\frac{\partial f(x)}{\partial x_m}\right|<\delta\right\}.
$$
Then
\begin{eqnarray*}
\|PU_{QFT}\ket{\psi_L}\|_2
& = & \left|e^{2\pi i\lambda(f(x)-\mu\nabla f(x)\cdot g_0)}\right|\prod_{m=1}^p\|P_m\ket{\phi_m}\|_2 \\
& = & \prod_{m=1}^p\sqrt{1-\|(I-P_m)\ket{\phi_m}\|_2^2} \\
& = & \prod_{m=1}^p\sqrt{1-\sum_{h_m=0}^{2^n-1}\left\{|\braket{h_m}{\phi_m}|^2\suchthat
 \left|c_{g,m}(h_m)-\frac{\partial f(x)}{\partial x_m}\right|\geq\delta\right\}} \\
& \geq & \prod_{m=1}^p\sqrt{1-2^n(2^{-n}|\csc(\pi\lambda\mu\delta)|)^2} \\
& = & (1-2^{-n}\csc^2(\pi\lambda\mu\delta))^{p/2} \\
& \geq & (2+\epsilon)/3,
\end{eqnarray*}
by~(\ref{ineqLeakage}).

By the triangle inequality,
$$
\|P\ket\chi\|_2 \geq \|PU_{QFT}\ket{\psi_L}\|_2 - \|PU_{QFT}\ket{\psi_D}\|_2 - \|PU_{QFT}\ket{\psi_D}\|_2.
$$
Since $U_{QFT}$ is an isometry and $P$ is a projection and therefore a contraction,
$$
\|P\ket\chi\|_2 \geq \frac{2+\epsilon}3 - \frac{1-\epsilon}3 - \frac{1-\epsilon}3 = \epsilon.
$$
\end{proof}

\section{Some implementation and efficiency considerations}  \label{secImplementation}

Theorem~\ref{thmCgnce} established that
the algorithm $A(n, \nu, \lambda, \mu; x)$
can perform at any required level of precision, given suitable operating parameters.
The algorithm consists of two quantum Fourier transforms, a phase rotation operator,
and the two operations $U_+$ and $U_f$ together with their inverses.

Because we have restricted the sampling grid side-length to powers of two,
the quantum Fourier transform is easily computed.
The standard quantum Fourier transform in a $2^n$-dimensional state space uses just $n$ Hadamard gates
and $n(n-1)/2$ controlled phase rotation gates; see~\cite{QCText} for details.
The $p$-dimensional quantum Fourier transform $U_{QFT}$ required by the gradient estimation algorithm
is simply the $p$th tensor power of the standard quantum Fourier transform,
meaning that it can be implemented by applying the standard quantum Fourier transform,
simultaneously but independently, to the $p$ factors of $\cal G$.

The difficulty of implementing the phase rotation operator $U_R$ depends on the data storage method used for function values,
i.e., on $({\cal R}, B_{\cal R}, c_r)$.  Implementation is straightforward if a binary fixed-point representation is used,
that is, if $\cal R$ is the state space of a system of say $N$ bits,
and the bit sequence
$$
(r_{N-1}, \ldots, r_1, r_0)
$$
represents the value $a_0+a_1\sum_{k=0}^{N-1}2^kr_k$, for some real constants $a_0$ and $a_1$.
In this case we can simply pass each bit independently through a phase rotation gate, with matrix representation
$$
\left(\begin{array}{cc}1 & 0 \\ 0 & e^{2\pi i\lambda a_12^k}\end{array}\right);
$$
these phase rotation gates are similar to, but simpler than,
the controlled phase rotation gates used in the quantum Fourier transform.

The operators $U_+$ and $U_f$ and their inverses simply perform machine arithmetic.
The operators $U_+$ and $U_+^{-1}$ each involve one multiplication and one addition per domain dimension.
The complexity of these operations in gate operations depends on the precision required.

The computational complexity of the operator $U_f$ is entirely dependent on the given function $f$.
It is usual in complexity analyses of computations involving a black-box function $f$ to assume that evaluations of $f$
will be the dominating cost, measuring complexity by counting function evaluations.
By that measure, the gradient estimation algorithm scores very well, as it requires two oracle operations,
$U_f$ and $U_f^{-1}$, the latter having presumably the same complexity as the former.
(In fact, recall that Section~\ref{secOracle} assumed $B_{\cal R}$ to be a group;
if this group is taken to be $\Integers_2^N$, that is, if the computed value is stored in $\cal R$ using the {\tt XOR} operation,
then $U_f^{-1}$ is just $U_f$.)

Of course, in order to perform at the required level of precision,
we may require very great accuracy in the evaluations of $f$, and in the other computations.
Note that this is a universal feature of machine computation.

\section{Conclusion}

Theorem~\ref{thmCgnce} shows that the gradient of a real-valued multivariate function can be evaluated
to any required accuracy using just two function evaluations.
As with any digital computation, increased accuracy in the answer requries increased precision during the computation.
Thus the quantum complexity of the gradient estimation problem is constant in dimension,
which compares favourably with the classical complexity, which is linear in dimension,
for very high-dimensional functions.

\def\refname{Reference}

\end{document}